\documentstyle[12pt,fullpage,epsf,psfig]{article}
\def\feyn{/\hspace{-.6em}}
\begin{document}
%\centerline{DRAFT \today \ 
%\tiny\verb$Id: note1.tex,v 1.15 1998/07/07 17:19:21 misha Exp $}
\hfill hep-ph/9804290\\
\vskip 1in
\centerline{\LARGE\bf 
On the Phase Diagram of QCD}
\vskip 1in
\centerline{\rm 
M.\,A. Halasz$^{\rm a}$, A.\,D. Jackson$^{\rm b}$, R.\,E. Shrock$^{\rm c}$,}
\centerline{\rm 
M.\,A. Stephanov$^{\rm c}$, and J.\,J.\,M. Verbaarschot$^{\rm a}$ }
\vskip 0.5in
\centerline{$^{\rm a}$ \it 
Department of Physics and Astronomy, SUNY, Stony Brook, NY 11794, USA}
\centerline{$^{\rm b}$ \it 
The Niels Bohr Institute, Blegdamsvej 17, Copenhagen, DK-2100, Denmark}
\centerline{$^{\rm c}$ \it 
Institute for Theoretical Physics, SUNY, Stony Brook, NY 11794-3840, USA}

\vskip 1.5in

\centerline{\large\bf Abstract}
\bigskip

We analyze the phase diagram of QCD with two massless quark flavors in the
space of temperature, $T$, and chemical potential of the baryon
charge, $\mu$, using available experimental knowledge of QCD, insights gained
from various models, as well as general and model independent
arguments including continuity, universality, and thermodynamic
relations. A random matrix model is used to describe the chiral symmetry 
restoration phase transition at finite $T$ and $\mu$.  In agreement with 
general arguments, this model predicts a tricritical point in the $T \mu$
plane.  Certain critical properties at such a point are universal and
can be relevant to heavy ion collision experiments.

\newpage

\section{Introduction}

Current and projected experimental progress in the physics of heavy ion 
collisions increasingly demands better theoretical understanding of the 
underlying phenomena.  In particular, the most exciting possibility offered 
by such experiments is the creation of high temperature and density
conditions under which the dynamics of QCD can bring matter into a new state. 
The challenge is then to calculate the properties of this new phase
together with the properties of the phase transition from QCD,
the underlying theory of quark-gluon interactions. 

Substantial progress has been achieved in our understanding of QCD at high 
temperature, $T$.  The foundation of this understanding is provided by 
lattice field theory Monte Carlo calculations.  In particular, we know that 
in QCD with two massless flavors a transition restoring chiral 
symmetry occurs at a temperature of approximately $160$ MeV \cite{TC93}.

On the other hand, little is known about the behavior of QCD for finite 
baryon charge density, or chemical potential of the baryon charge,
$\mu$.  Standard lattice Monte Carlo techniques cannot be applied since 
the determinant of the Dirac operator is complex, and hence the
Euclidean path integral
defining the theory does not have a Gibbs (i.e., real, positive-definite)
measure. A Gibbs measure is needed for the probabilistic
interpretation which forms the basis for importance
sampling methods such as Monte Carlo calculations. Moreover, the 
approximation of quenched 
fermions fails in this case \cite{Ba86,KoLo95} for reasons which
have been understood recently using the random matrix theory \cite{St96b}.  
However, the conditions created in heavy ion collision experiments require 
an understanding of the regime of high baryon density as well as 
that of high temperature.   As reviewed, e.g., in \cite{stachel96}, there
is good evidence that central part of the collisions 
can be described approximately before freeze-out by 
thermodynamics, so that the 
temperature and chemical potential can be defined. 

The purpose of this paper is to assemble available knowledge about
QCD and apply it to the construction of the phase diagram in the $T \mu$
plane.  Most of the studies of this phase diagram have concentrated on 
modeling the properties of the chiral phase transition (see, e.g., 
\cite{AsYa89,Ba89-94,BiDe91,Kl92,BrBu96}). In this paper, we present a more complete and less 
model dependent analysis of the phase diagram which also includes effects 
from other phase transitions, such as the nuclear matter liquid-gas 
transition.  Naturally, many of the phenomena to be discussed have been 
studied extensively.  As a result, we will repeat some familiar 
experimental facts and theoretical arguments (with references to some of 
the original papers or reviews as appropriate).  The aim of our analysis 
is to transform 
this knowledge into the determination of a phase diagram for QCD in the $T \mu$ plane.  Such an analysis is especially important as an extension of Monte
Carlo studies, given the technical problems that these
encounter with finite baryon charge density.

The chiral phase transition is of primary interest in ultrarelativistic 
heavy ion experiments since this is the transition that separates the hadronic 
phase from the quark-gluon phase.  In section~\ref{sec:rmt}, we introduce a 
random matrix model of the chiral phase transition at finite $T$ and $\mu$.  
We find that this model predicts a tricritical point in the $T \mu$ plane 
in agreement with more generic arguments.  We analyze the properties of 
some thermodynamic observables in the vicinity of this point.

\section{Definitions}

We take as our model the standard approximation in which we (i) consider 
pure SU(3) QCD with electroweak interactions turned off and (ii) consider 
this theory with two massless quarks.  There is then an exact
SU(2)$_L\times$SU(2)$_R\times$U(1)$_B$ global symmetry of 
the action, which is 
spontaneously broken down to SU(2)$_V\times$U(1)$_B$ at zero and 
sufficiently low temperatures by the formation of a condensate,
$\langle\bar\psi\psi\rangle$.  Many 
features of QCD indicate that this is a reasonable approximation, e.g., 
the lightness of pions, the success of current algebra relations, etc. 
(We will comment below on the inclusion of electromagnetic interactions 
and strange quarks.)  This theory is described by a grand canonical
partition function 
which, when written as a path integral, is formally:
\begin{equation}\label{z}
Z \equiv e^{-\Omega(T,\mu)/T} = \int {\cal D}A {\cal D}\bar\psi {\cal D}\psi
\exp\{-S_E\} \ .
\end{equation}
The Euclidean action, $S_E$, is given by
\begin{equation}
S_E = \int_0^{1/T} dx_0 \int d^3x 
\left[ {1\over2g^2} {\rm Tr} F_{\mu\nu}F_{\mu\nu} 
- \sum_{f=1}^{N_f} \bar\psi_f\left(\feyn\partial + \feyn{A}
+ m_f + {\mu\over N_c}
\gamma_0\right)\psi_f
\right] \ ,
\end{equation}
where $N_f=2$ is the number of flavors, $N_c=3$ is the number of
colors, and $m_f=m=0$ is the quark mass.
The Euclidean matrices $\gamma_\mu$ are hermitean.
Note that with our sign choices positive $m$ and $\mu$
induce positive $\langle\bar\psi\psi\rangle$
and $\langle\bar\psi\gamma_0\psi\rangle$.
The normalization of $\mu$ differs from the normalization 
customary in lattice calculations by a factor $1/N_c$ (i.e., the baryon 
charge of a quark).
Integrating over the fermion fields we can also write:
\begin{equation}
Z = \int {\cal D}A \exp\left[-{1\over2g^2} 
{\rm Tr} F_{\mu\nu}F_{\mu\nu}\right]
\det\left[\feyn{D} + m_f + {\mu\over N_c}\gamma_0\right].
\end{equation}

As indicated, this system is characterized by equilibrium values of $T$ and
$\mu$. This may be thought of by imagining the system to be in 
thermodynamic equilibrium with a large reservoir of entropy and baryon 
charge which is characterized by these values of $T$ and $\mu$.
The total energy and 
baryon charge of our system fluctuate.  Of course, the relative 
magnitude of these fluctuations is negligible for an open system of
macroscopic size. 
The relation between the chemical potential, $\mu$, and the average baryon 
number density (per unit volume), $n$, is the same as that between the 
temperature, $T$, and the average entropy density (per unit volume), $s$:
\begin{equation}\label{ns}
nV= \sum_f\langle\bar\psi_f\gamma_0\psi_f \rangle
= -{\partial\Omega\over \partial\mu}\ ; \qquad
sV = -{\partial\Omega\over \partial T}\ ,
\end{equation}
where $\Omega$ is the thermodynamic potential defined in eqn.\,(\ref{z}). 
It can also be seen that $\Omega=-pV$, where $p$ is the pressure.
In other words, pressure, temperature and chemical potential
are not independent variables for our system. Their variations
are related by
\begin{equation}\label{dp}
 dp = sdT + nd\mu. 
\end{equation}

Both $T$ and $\mu$ (as well as $p$) are intensive parameters.  For a 
system in thermodynamic equilibrium, these quantities are the same for 
any of its smaller subsystems.  In contrast, the extensive 
 densities $s$ and $n$ 
can differ for two subsystems even when they are in equilibrium with each
other.  This happens in the phase coexistence region, e.g., a glass 
containing water and ice.  It is more convenient to describe the phase
diagram in the space of intensive parameters, $T$ and $\mu$.  In
particular, the first-order phase transition which we shall encounter is
characterized by one value of $\mu$ but two values of $n$ --- the
densities of the two coexisting phases.  Another reason for working in
these coordinates is that first-principle lattice calculations are 
performed in such a way that $T$ and $\mu$ are the parameters that 
can be controlled while the densities are measured.  The results of 
relativistic heavy ion collision experiments are also often analyzed 
using this set of parameters \cite{stachel96}.

\section{Zero Temperature}

We begin by considering the phase diagram as $\mu$ is varied along the 
line $T=0$.  Strictly speaking, we are not dealing with
{\em thermo\/}dynamics here since the system is in its ground state.
This fact leads to a simple property of the function $n(\mu)$.  Let 
us rewrite the partition function, eqn.\,(\ref{z}), as the Gibbs sum over 
all quantum states, $\alpha$, of the system:
\begin{equation}\label{gibbs}
Z = \sum_\alpha \exp\left\{-{E_\alpha - \mu N_\alpha\over T}\right\}\ ,
\end{equation}
where each state is characterized by its energy, $E_\alpha$, and
its baryon charge, $N_\alpha$. In the limit $T \to 0$, the state with 
the lowest value of $E_\alpha-\mu N_\alpha$ makes an exponentially 
dominant contribution to the partition function.  When $\mu=0$, this is 
the state with $N=0$ and $E=0$, i.e., the vacuum or $\alpha=0$.  Let us 
introduce 
\begin{equation}\label{mu0def}
\mu_0 \equiv \min_\alpha (E_\alpha/N_\alpha) \ .
\end{equation}
As long as 
$\mu < \mu_0$, the state with the lowest of $E_\alpha-\mu N_\alpha$ 
remains the vacuum, $\alpha=0$.  Therefore, we conclude that at zero
temperature
\begin{equation}\label{n0}
n(\mu) = 0 \quad {\rm for} \quad \mu < \mu_0 \ .
\end{equation}

What is the value of $\mu_0$?  As an exercise, we first consider a free 
theory of massive fermions carrying one unit of baryon charge. The 
states which minimize $E_\alpha / N_\alpha$ are states with one or two 
(more if fermions have flavor or other degeneracy) fermions at rest with 
$p=0$.  For each of these states, $E_\alpha / N_\alpha = m$, the mass of 
the fermion.  Therefore, $\mu_0=m$ for such a theory.  When $\mu > m$, 
the ground state is the Fermi sphere with radius $p_F=\sqrt{\mu^2-m^2}$. 
Therefore, $n(\mu)=(\mu^2-m^2)^{3/2}/(3\pi^2)$.  Thus, we see that, even 
in a trivial theory, the function $n(\mu)$ has a singularity at 
$\mu = \mu_0$.  The existence of some singularity at the point $\mu=\mu_0$, 
$T=0$ is a robust and model independent prediction. This follows from the 
fact that a singularity must separate two phases distinguished by an order 
parameter, e.g., $n$.  The function $n \equiv 0$ cannot be continued to 
$n \not= 0$ without a singularity. 

What is $\mu_0$ for the case of QCD, and what is the form of the singularity? 
The answers to these questions are somewhat different in QCD and in the 
real world (QCD+) which includes other interactions, most notably 
electromagnetic interactions.  Since QCD is the focus of the present 
paper and QCD+ is the ultimate goal of our understanding, we shall consider
both cases.  It is important to understand their differences if we are 
to extract physically useful predictions from lattice calculations, which 
are performed for QCD rather than QCD+.

\begin{figure}
\centerline{\epsfysize 2in\epsfbox{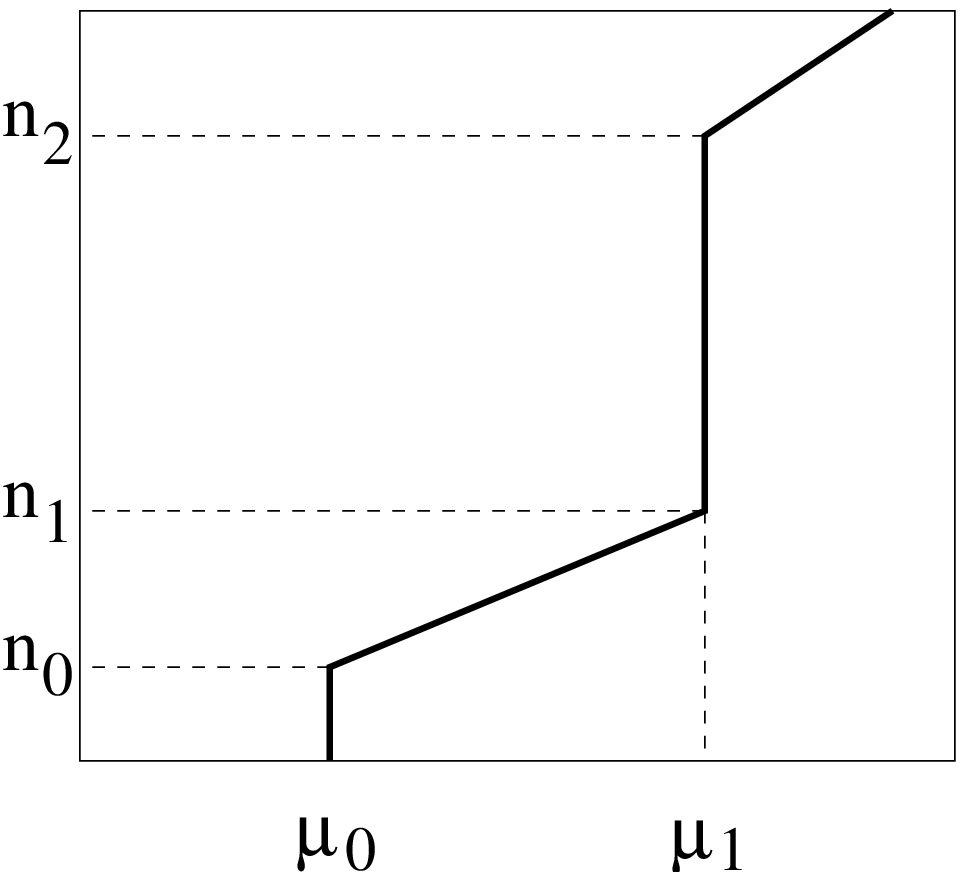}\qquad\qquad
\epsfysize 2in\epsfbox{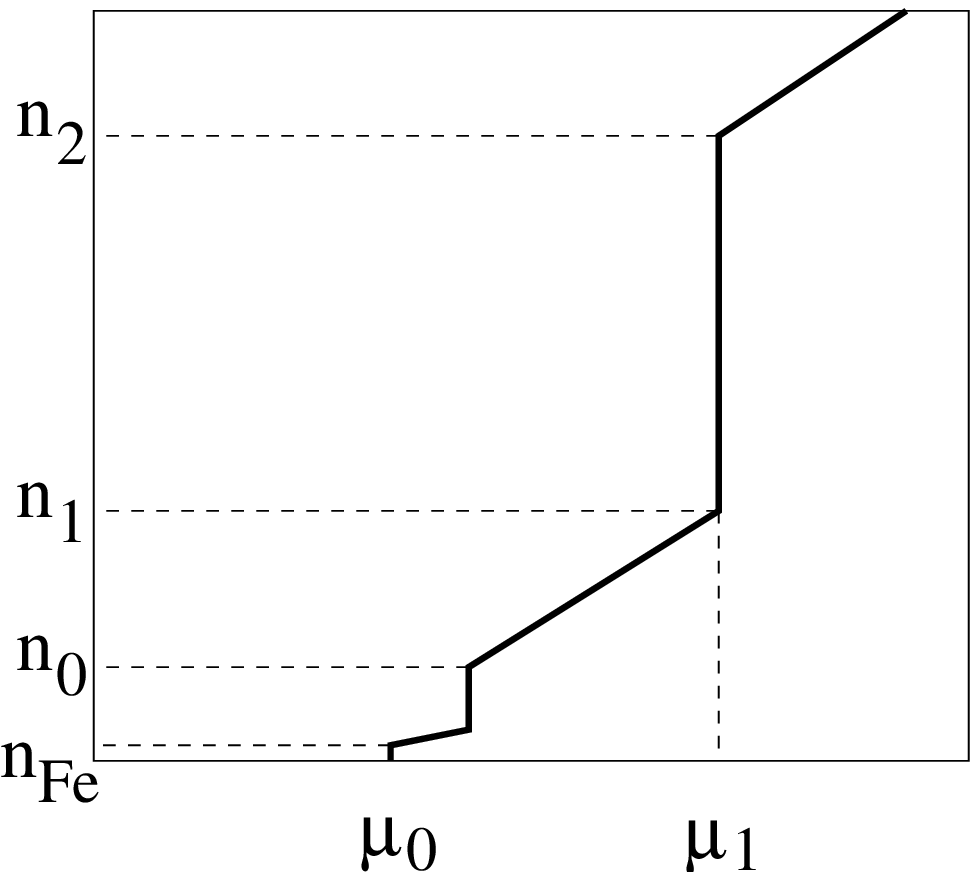}\\}
\centerline{\hss (a) \hss (b) \hss}
\caption[]{\small Schematic dependence of the baryon charge density
on the chemical potential at $T=0$ (a) in QCD ($\mu_0\approx m_N-16$ MeV) 
and (b) in QCD+ ($\mu_0\approx m_N-8$ MeV).}
\label{fig:nmu}
\end{figure}

The energy per baryon, $E/N$, can also be written as
$m_N - (Nm_N-E)/N$, where $m_N=m_p\approx m_n$ is the nucleon mass. 
Therefore, the state which minimizes
$E/N$ is that for which the binding energy per nucleon,
$\epsilon=(N m_N - E)/N$, is a maximum.
Empirically, we know that this state is a single
iron nucleus at rest with $N=A=56$ and $\epsilon\approx 8$ MeV.
However, in QCD without electromagnetism the binding energy per nucleon 
increases with $A$. This is the consequence of the saturation of
nuclear forces and can be seen from the Weizsacker
formula. Without electromagnetism, only the bulk and surface energy
terms are significant for large  $A$:
\begin{equation}
\epsilon(A) \equiv {Am_N-m_A\over A}\approx a_1 - a_2A^{-1/3}
\end{equation}
with $a_1\approx 16$ MeV, $a_2\approx 18$ MeV \cite{FeWa}.  As $A \to 
\infty$, $\epsilon$ saturates at the value $a_1$. This corresponds
to the binding energy per nucleon in a macroscopically large sample of
nuclear matter as defined by Fetter and Walecka in \cite{FeWa}.
We conclude that in QCD the density jumps at $\mu=\mu_0\approx m_N
-16$ MeV to the value of the nuclear matter density $n_0\approx
0.16\,{\rm fm}^{-3}$. Therefore, in QCD there is a first-order
phase transition, characterized by a discontinuity in the function
$n(\mu)$ at $\mu=\mu_0$ (see Fig.~\ref{fig:nmu}a).

In QCD+, the Coulomb forces change the situation near $\mu_0$. 
The contribution of the Coulomb repulsion to $\epsilon(A)$ is
negative: $- (0.7\,{\rm MeV})\,Z^2/A^{2/3}$, and it is responsible 
for the experimentally observed maximum in $\epsilon(A)$ at $A \approx 56$. 
  Isospin singlet nuclear matter ($A = \infty$) is unstable at zero
pressure due to Coulomb 
repulsion.  Neutron matter with $Z \ll A$ is also unstable at zero
pressure, and we are left to consider a gas of iron nuclei. In order to ensure
electric neutrality, we must add electrons.  Such a gas is clearly
unstable at small densities and forms a solid --- iron.  Therefore, there 
is a discontinuity in the value of $n(\mu)$ at $\mu_0\approx m_N - 8$ MeV.  
This discontinuity is equal to the density of normal matter (i.e., iron) 
and is about $10^{-14}$ times smaller than in QCD.  
For very small $\mu-\mu_0$, $n(\mu)$ 
has structure, fine on the scale of QCD, which reflects the properties of
normal matter under pressure.  Then, for $\mu - \mu_0 = {\cal O}(10-200\,
{\rm MeV})$, we traverse the domain of nuclear physics with the possibility 
for various phase transitions.  In particular, a transition to neutron 
matter ($Z \ll A$) is probably similar to the transition in QCD at 
$\mu = \mu_0$.  (See Fig.~\ref{fig:nmu}b.)  In this domain, one may encounter 
such phenomena as nuclear matter crystallization \cite{Mi90,Kl85}, 
superconducting phases of neutron and quark matter 
\cite{BaLo84,RaSc97,AlRa97}, 
and, due to the strange quark in QCD+, kaon condensation \cite{Mi90,KaNe86}
and a transition to strange quark matter \cite{Wi84,FaJa84}. 
Moving along the $\mu$ axis to the right is equivalent to increasing the 
pressure: $p=\int n d\mu$.  Thus, this picture is roughly what one might 
encounter in moving towards the center of a neutron star from 
the iron crust at the surface.

Our knowledge of $n(\mu)$ is scanty for densities of order one to ten times 
$n_0$ and  $\mu - \mu_0 = {\cal O}(10-200\,{\rm MeV})$ both in QCD and in 
QCD+.  We can only be sure that $n(\mu)$ is a monotonically increasing 
function, which follows from the requirement of thermodynamic stability.

The behavior of $n(\mu)$ again becomes calculable in the region of very 
large $\mu \gg \Lambda_{QCD}$.  In that case, the Pauli exclusion principle
forces the quarks to occupy ever higher momentum states, and, due to 
asymptotic freedom, the interaction of quarks near the Fermi surface is 
(logarithmically) weak.  The baryon charge density is proportional to the 
volume of a Fermi sphere of radius $\mu/3$, $n(\mu) \approx N_f (\mu/3)^3 
/ (3 \pi^2)$.  At low temperatures, only quarks near the Fermi surface 
contribute to the Debye screening of the gauge fields.  The square of the 
screening mass, $m_D^2$, is proportional to the area of the Fermi surface:
$m_D^2\sim g^2\mu^2$.  This means that color interactions are screened
on lengths ${\cal O}(1/g\mu) = {\cal O}(\sqrt{\ln(\mu/\Lambda_{\rm QCD})} 
/\mu)$.  This motivates the conclusion that nonperturbative phenomena such 
as chiral symmetry breaking should be absent at sufficiently large
$\mu$.  Therefore, in QCD with massless quarks one should expect at least
one other phase transition, at a value of $\mu$ which we define as 
$\mu_1$ --- a transition 
characterized by the restoration of the chiral symmetry.

What is the value of $\mu_1$ in QCD, and is it finite?
Very little reliable information about the phase transition at $\mu_1$
is available. However, several different approaches agree on the conclusion 
that the value of $\mu_1$ is finite and that $\mu_1 - \mu_0$ is on the order 
of the typical QCD scale $\Lambda_{\rm QCD}\approx 200\,{\rm MeV} \approx 
1\,{\rm fm}^{-1}$.  For example, equating the quark pressure minus the MIT 
bag constant to the pressure of nuclear matter yields such an estimate 
(see, e.g., \cite{CsKa86}). 
Here, we should also point out another interesting distinction between 
QCD and QCD+: the effect of the strange quark in QCD+ is to decrease the 
value of $\mu_1$ compared to that of QCD.  It has even been conjectured 
that this effect might be sufficient to drive $\mu_1$ below 
$\mu_0$, which would make normal nuclear matter metastable \cite{Wi84,FaJa84}.
Another model which predicts the phase transition at finite $\mu_1$ is the 
Nambu-Jona-Lasinio model, which focuses on the degrees of freedom associated 
with the spontaneous chiral symmetry breaking and leads to a similar 
estimate for $\mu_1$ \cite{Kl92}.

What is the order of this phase transition? The MIT bag model predicts
that it is a first-order transition since the density, $n$, 
of the baryon charge
is discontinuous.  Unfortunately, analysis of the Nambu-Jona-Lasinio 
model shows that the order of the transition depends on the values
of parameters, most notably, on the value of the cutoff.  A larger cutoff 
leads to a second-order transition, a smaller to a first-order transition 
\cite{Kl92}.  A random matrix model at $T=0$ predicts a first-order phase 
transition \cite{St96b}.  In this paper, we shall extend the random matrix 
model to permit consideration of the entire $T \mu$ plane.  Before doing 
this, we shall use more general methods to analyze features of the phase 
diagram of QCD at finite density {\em and\/} temperature in the next section.

An additional, qualitative argument for the first-order nature of the
chiral phase transition at $\mu_1$ can be also drawn from a certain
analogy of QCD to a metamagnet such as a crystal of ferrous chloride
FeCl$_2$.   At temperatures below the N\'eel temperature, $T_N$, and at zero
magnetic field, $H$, such a
crystal is antiferromagnetically ordered (i.e., the staggered 
magnetization, $\phi_{\rm st}$, has a nonzero expectation value: 
$\langle \phi_{\rm st}\rangle\not=0$.
Analogously, $\langle\bar\psi\psi\rangle \not= 0$ in QCD below $T_c$.
The magnetic field $H$ is not an ordering field
for the staggered magnetization because it couples to a 
different order parameter (i.e., normal magnetization, 
$\phi$, with $\Delta {\cal E} =
-H\phi$) and induces nonzero $\langle\phi\rangle$. 
Similarly, the chemical potential induces nonzero
$\langle\bar\psi\gamma_0\psi\rangle$, and the term 
$\mu\bar\psi\gamma_0\psi$ does not introduce explicit breaking
of the chiral symmetry. At some critical value of $H$, 
ferrous chloride undergoes a first-order phase transition, and the
staggered magnetization vanishes: $\langle\phi_{\rm st}\rangle = 0$. One
could naturally expect that in QCD a similar competition between the
low temperature spontaneous ordering, $\langle\bar\psi\psi\rangle
\not= 0$, and the ordering $\langle\bar\psi\gamma_0\psi\rangle\not=0$ induced
by $\mu$ would result in a first-order phase transition. This analogy can
be continued into the $T \mu$ plane or the $T H$ plane in the case of the
antiferromagnet. The antiferromagnet has a well known tricritical
point in this plane. Its analogue in QCD will be discussed in
section~\ref{sec:3cr}.

Following the arguments of the two preceding paragraphs, we base our 
subsequent analysis of the phase diagram of QCD with two massless quarks
on the following expectations: (i) $\mu_1 \sim \mu_0 + {\cal O}(200 \, 
{\rm MeV})$ and (ii) the transition is of first order. 

\section{Finite $T$ and $\mu$}

We shall use two order parameters to analyze the phase diagram
of QCD at nonzero $T$ and $\mu$: the chiral condensate
$\langle\bar\psi\psi\rangle$ (per flavor) given by
\begin{equation}
\langle\bar\psi\psi\rangle V = -{1\over N_f}{\partial\Omega\over \partial m}
\end{equation}
and the density of the baryon charge $n$ given by eqn.\,(\ref{ns}). 
  We have 
already used $n$ to show that there is a singularity at $\mu = \mu_0$ and 
$T=0$.  It was important for that argument that $n$ is exactly zero for all 
$\mu<\mu_0$.  At nonzero $T$, however, $n$ is not strictly $0$ for any 
$\mu > 0$.  For example, for very small $\mu$ and $T$ one finds a very dilute
gas of light mesons, nucleons and antinucleons with
\begin{equation}
n(T,\mu)\approx {\mu\over T} \left(2m_NT\over\pi\right)^{3/2} e^{-m_N/T} \ .
\end{equation}

Nevertheless, we can use a continuity argument to deduce that
the first-order phase transition at $T=0$, $\mu=\mu_0$ has to
remain a first-order phase transition for sufficiently small $T$.
  Therefore, there must be a line 
emerging from the point  $T=0$, $\mu=\mu_0$.  One can think of this 
transition as boiling the nuclear fluid.  The slope of this line can 
be related to the discontinuities in the entropy density, $\Delta s$ 
(or the latent heat per volume $T\Delta s$), and in the baryon density, $\Delta n$, 
across the phase transition line through the generalized 
Clapeyron-Clausius relation:
\begin{equation}\label{CC}
{dT\over d\mu} = -{\Delta n\over \Delta s} \ .
\end{equation}
This relation follows from the condition that the pressure,
 temperature and chemical potential should be
the same in the two phases on a phase coexistence curve and
eqn.\,(\ref{dp}).   In analogy with ordinary 
liquid-gas transitions, the gaseous phase has a lower particle density 
(whence $\Delta n < 0$) and lower entropy density%
\footnote{The entropy per particle is greater in the gaseous phase,
but the entropy per volume $s$ is smaller because of much smaller
particle density.} 
(whence $\Delta s < 0$).  Therefore, 
the slope $dT/d\mu$ must be negative.  We further expect that the slope is 
infinite at $T=0$ since $s(T=0)=0$, and hence $\Delta s(T=0)=0$.   
As there is no symmetry-breaking order 
parameter which distinguishes the two phases, there is no reason why these 
two phases cannot be connected analytically.  As in a typical liquid-gas 
transition, it is natural to expect that the first-order phase transition 
line terminates at a critical point with the critical exponents of the 
three-dimensional Ising model.%
\footnote{ The Ising nature of the universality class
follows from the fact that the transition can be modeled by an Ising lattice
gas.}
 The temperature of this critical point
can be estimated from the binding energy per nucleon in cold nuclear matter, 
$T_0= {\cal O}(10 \, {\rm MeV})$. (See Fig.~\ref{fig:pd}.)  Signatures of 
this point are seen in heavy ion collisions at moderate energies (i.e., 
$\approx 1$ GeV per nucleon), and the critical properties of this point have 
been studied through measurements of the yields of nuclear fragments 
\cite{CsKa86,Tr96}.  In particular, the reported critical exponents are in 
agreement with those of the three-dimensional Ising model \cite{Tr96}.

Additional phase transitions which might occur at $T=0$ would give rise to 
additional phase transition lines.  One could expect two generic situations.
If there is a breaking of a global symmetry (e.g., translational symmetry in 
the case of nuclear matter crystallization), the phase transition line must 
separate such a phase from the symmetric phase at higher temperature without 
any gaps in the line.  Otherwise, the transition can terminate at a critical 
point.

At very high $T \gg \Lambda_{\rm QCD}$, we have a plasma of quarks and
gluons with a logarithmically small effective coupling constant, $g(T)$,  
and we can again calculate the density of the baryon charge $n$:
\begin{equation}
n(T,\mu)\approx 4\int {d^3p\over(2\pi)^3} \left[
\exp{|\mbox{\boldmath $p$}|-\mu/3\over T}+1
\right]^{-1} - \ \{ \mu \to -\mu \} \ .
\end{equation}
We expect that the chiral condensate is zero at very high $T$ since
the effective coupling is weak because of asymptotic freedom.  
Therefore, a phase transition must 
separate the quark gluon phase from the low temperature phase. This 
transition has been studied extensively at $\mu=0$ using a variety of 
methods.  In particular, lattice calculations have established the value 
of $T_c$ as approximately $160$ MeV \cite{TC93}.  Arguments based on 
universality suggest that this transition is of second order with 
critical exponents of the SU(2)$_L \times$ SU(2)$_R \sim$ O(4)
universality class \cite{PiWi84}.  Lattice calculations seem to
confirm this scenario \cite{Be97}.\footnote{A sufficiently light third quark
would drive the transition first-order \cite{PiWi84}.  However, lattice 
calculations also indicate that the strange quark is not sufficiently 
light for this to occur \cite{columb90}.}  Here, we assume that this
is the case and try to understand what happens to this transition when 
$\mu$ is not zero.

For massless quarks, the low--temperature hadronic phase and the
quark-gluon plasma phase can be distinguished by the expectation
value of $\langle {\bar \psi} \psi \rangle$, since this is identically zero in 
the quark-gluon phase and nonzero in the hadronic phase with 
spontaneously broken chiral symmetry.  Therefore, when quark masses are 
strictly zero, a phase transition must separate these two phases, i.e., 
these phases cannot 
be connected analytically in the $T \mu$ plane at $m=0$. Therefore, a line
of phase transitions must begin from the point $T=T_c$, $\mu=0$ 
and continue into the $T \mu$ plane.

As discussed above, chiral symmetry restoration
at $T=0$ is most likely to proceed via a first-order phase transition.
Therefore, the transition must remain first-order as we continue along 
a line into the $T \mu$ plane.  The slope of this line can again be related 
to the discontinuity in the baryon charge and the entropy density 
(\ref{CC}).  Since we expect that both density and entropy  will be larger 
in the quark-gluon phase, the slope of this line, $dT/d\mu$, should be 
negative.  

This first-order transition line cannot terminate 
because the order parameter, $\langle {\bar \psi} \psi \rangle$, is
identically zero on the one side of the transition. The minimal
possibility is that it merges with the second-order phase transition line
coming from $T=T_c$, $\mu=0$; the point where the two lines join is a
tricritical point \cite{AsYa89,Ba89-94,Kl92}.  Such a point exists in many physical systems
(e.g., in the FeCl$_2$ antiferromagnet), and universal behavior in the 
vicinity of this point has been studied extensively. In the next section, 
we review those properties of a tricritical point which follow from 
universality.

\section{Universal properties of the tricritical point}
\label{sec:3cr}

By analogy with an ordinary (bi)critical point, 
where two distinct coexisting phases become identical, one can define 
the tricritical point as a point where three coexisting phases become 
identical simultaneously.  A tricritical point marks an end-point 
of three-phase coexistence. In order to see this in QCD, it is necessary 
to consider another dimension in the space of parameters --- the quark 
mass $m$.  This parameter breaks chiral symmetry explicitly.  In 
such a three-dimensional space of parameters, one can see that there are 
two surfaces (symmetric with respect to $m\to-m$ reflection) of first-order 
phase transitions emanating from the first-order line at $m=0$.  On these 
surfaces or wings with $m \not= 0$, two phases coexist: a low density phase 
and a high density phase.  There is no symmetry distinguishing these two 
phases since chiral symmetry is explicitly broken when $m \ne 0$. Therefore,
the surfaces can have an edge which is a line of critical points. These 
lines, or wing lines, emanate from the tricritical point. The first-order 
phase transition line can now be recognized as a line where three phases 
coexist: the high $T$ and density phase and two low density and $T$ phases 
with opposite signs of $m$ and, hence, also of $\langle {\bar \psi} 
\psi\rangle$.  This line is called, therefore, a triple line.

The plane $m=0$ is a symmetry plane.  Chiral symmetry is exact only in 
this plane, and it is only here that the low and the high temperature 
phases must be separated by a transition.  One can also view this plane 
as a first-order phase transition surface, since $\langle {\bar \psi} 
\psi \rangle$ has a discontinuity across it.  Then, the second-order phase 
transition line together with the triple line provide 
a boundary for this surface.

Critical behavior near the tricritical point can now be inferred from 
universality.  The upper critical dimension for this point is 3. Since 
critical fluctuations are effectively three-dimensional for the 
second-order phase transition at finite $T$, we conclude that behavior 
near this point is described by mean field exponents with only logarithmic 
corrections.  The effective Landau-Ginsburg theory for the long-wavelength
modes, $\phi \sim \langle {\bar \psi} \psi \rangle$, near this point 
requires a $\phi^6$ potential which has 
the form (in the symmetry plane $m=0$)
\begin{equation}\label{LG}
\Omega_{\rm eff} = \Omega_0(T,\mu) + a(T,\mu)\phi^2 + b(T,\mu)\phi^4 + 
c(T,\mu)\phi^6
\end{equation}
with $c > 0$. The $\phi^6$ term is necessary in order to create three 
minima corresponding to the three coexisting phases.  This explains why 
the critical dimensionality is $3$, since for this dimension, the operator 
$\phi^6$ becomes a marginal operator.  When $b > 0$, the transition occurs 
when $a = 0$ and is a second-order transition similar to that seen in a 
$\phi^4$ theory.  This corresponds to the second-order line.  
When $b < 0$ the transition 
occurs at some positive value of $a$ and is of first order.  This is 
the triple line. When both $a$ and $b$ vanish, we have a tricritical point. 

In particular, the following exponents in the symmetry plane $m=0$
are readily found using mean field $\phi^6$ theory (as noted above,
renormalization group studies \cite{LaSa} show that the actual singularities
include additional, logarithmic corrections).  The discontinuity 
in the order parameter  $\langle \phi \rangle = \langle {\bar \psi} \psi 
\rangle$ along the triple line as a function of the distance from the 
critical point $\mu_3$, $T_3$ (measured either as $T_3-T$ or $\mu-\mu_3$) 
behaves like
\begin{equation}\label{deltapbp}
\Delta \langle {\bar \psi} \psi \rangle \sim (\mu-\mu_3)^{1/2} \ .
\end{equation}
The discontinuity in the density, $n = d\Omega_{\rm eff}/d\mu$, across 
the triple line behaves like
\begin{equation}\label{deltan}
\Delta n \sim (\mu-\mu_3)^1.
\end{equation}

The critical behavior along the second-order 
line is everywhere the same as 
at the point $\mu=0$, $T=T_c$ (which is an infrared attractive fixed point).
Therefore, $\langle {\bar \psi} \psi \rangle$ vanishes on the second-order line
with $O(4)$ exponents.  At the tricritical point, however,
the exponent with which $\langle\bar\psi\psi\rangle$ vanishes
is given by the Landau-Ginzburg theory as
\begin{equation}\label{pbp1/4}
\langle\bar\psi\psi\rangle \sim (T_3-T)^{1/4}.
\end{equation}

When $m\not=0$, the potential $\Omega_{\rm eff}(\phi)$ can also contain
terms $\phi$ and $\phi^3$ which break $\phi\to-\phi$ symmetry explicitly.  
(The term $\phi^5$ can be absorbed by a shift of $\phi$.)  The
potential $\Omega_{\rm eff}(\phi)$ still has three minima, and a 
first-order phase transition can occur when two adjacent minima are equally
deep. These transitions form a surface of first-order phase
transitions --- the wings.  The two minima (and an intermediate maximum) 
can also fuse into a single minimum.  This happens on the wing lines at 
the edge of the surface of first-order phase transitions.  The critical 
behavior along the wing lines is given by the three-dimensional Ising 
exponents, as is usual at the endpoints of first-order liquid-gas type
phase transitions not associated with restoration of a symmetry.  
In particular, 
the discontinuity in $\Delta \langle {\bar \psi}\psi \rangle$ and $\Delta n$ 
vanishes with exponent $\beta \approx 0.31$.
 These discontinuities are related to the 
slope of the wing surface at constant $T$ through a relation similar
to (\ref{CC}):
\begin{equation}
{d\mu\over dm} = - {\Delta\langle\bar\psi\psi\rangle N_f\over\Delta n}.
\end{equation}

There are many other universal properties in the vicinity of a tricritical 
point which can be derived from the above $\phi^6$ Landau-Ginzburg effective 
potential.  One can, for example, show that the $m=0$
second-order line, the wing lines, and the triple 
lines approach the triple point with the same tangential direction: The 
second-order line approaches from one side while the wing lines and the triple line
approach from the opposite side.  For more detailed description of the 
properties of tricritical points, see ref.\,\cite{LaSa}.

\section{A random matrix model at finite $T$ and $\mu$}
\label{sec:rmt}

Random matrix models have proven to be a valuable tool for studying
spontaneous chiral symmetry breaking in QCD.  For example, it has been
conjectured that the distribution of the eigenvalues of the Dirac
operator near zero is universal \cite{ShVe93}.  The universal
expressions show a remarkable agreement with lattice Monte Carlo data
\cite{BeMe98} and are consistent with spectral sum rules from chiral
perturbation theory \cite{ShVe93}.
Random matrix models have also been used to study chiral symmetry 
restoration phenomenon at finite temperature
\cite{JaVe96,St96a,WeSc96} 
as well as finite chemical potential \cite{St96b,HaJa97,FeZe97,JaNo97}.

Random matrix theory provides an effective description of those degrees 
of freedom in QCD which are responsible for the spontaneous  breaking
of chiral symmetry. In this respect, it is similar to
Landau-Ginzburg effective theory. 
Random matrix theory is based on the observation that the spontaneous
chiral symmetry breaking is related to
the density of small eigenvalues of the Dirac operator ($\lambda \ll 
\Lambda_{\rm QCD}$).  This relationship is expressed quantitatively
by the Banks-Casher formula \cite{BaCa80}, 
$\langle {\bar \psi} \psi \rangle =\pi 
\rho_{\rm ev}(0)$.  Here, $\rho_{\rm ev}(0)$ is the density of small 
(but non-zero) eigenvalues (per unit $\lambda$ and per unit four-volume, 
$V_4$) of the Euclidean Dirac operator in 
the thermodynamic limit $V_4 \to \infty$.  The dynamics of these 
eigenvalues can be described using a random matrix (of infinite size) in 
place of the Dirac operator.  This approximation can be shown to give 
exact results in the mesoscopic limit \cite{ShVe93,JaSe96}.

When the chemical potential is non-zero, the Dirac operator is not 
hermitian, and its determinant is no longer real.  As a 
result, the density of its eigenvalues can be defined straightforwardly
only in quenched QCD, i.e., when the contribution of the (complex) 
fermion determinant to the measure is approximated by unity.  Fortunately, 
a more general relation exists between the chiral condensate and the linear 
density of zeroes of the partition function:
\begin{equation}\label{YL}
\langle\bar\psi\psi\rangle=\pi\rho(0) \ .
\end{equation}
 As observed by Yang and Lee \cite{YaLe52}, non-analytic behavior in a
thermodynamic quantity, including in the present case the discontinuity in 
the value of an order parameter in the thermodynamic limit, is caused by the
coalescence of zeros of the partition function to form a boundary crossing the
relevant parameter axis.  In the case of QCD, the signature of spontaneous 
chiral symmetry
breaking is the discontinuity in $\langle {\bar \psi} \psi \rangle$ as $m$ 
is varied along the real axis and crosses $m=0$.  This discontinuity is equal 
to $2\pi\rho(0)$ where $\rho(0)$ is the density of the zeroes on the 
imaginary $m$ axis near $m=0$ in the thermodynamic limit.  One can also show 
that $\rho \to \rho_{\rm ev}$ in the quenched limit $N_f \to 0$ but {\em 
only\/} when $\mu=0$.  For nonzero $\mu$, the density of the eigenvalues, 
$\rho_{\rm ev}$, is fundamentally different from the $N_f \to 0$ limit of 
$\rho$ \cite{St96b}.

It can be shown that certain properties of the small eigenvalues
 of the Dirac operator are universal and are identical 
to those of a random matrix model with appropriate symmetries
\cite{BrHi96,AkDa97,GuWe97,JaSe96,BeMe98,FyTi98}.
The virtue of the random matrix model is that it is solvable,
i.e., one can calculate the distribution of the eigenvalues and
of the Yang-Lee zeroes.
The partition function of QCD at finite temperature and chemical
potential in random matrix approximation is given by
\begin{equation}\label{ZRM}
Z_{\rm RM} = \int {\cal D}X \exp\left(-{N\over\sigma^2}{\rm Tr}XX^\dag\right) 
{\det}^{N_f}(D + m) \ ,
\end{equation}
where $D$ is the $2N\times 2N$ matrix approximating the
Dirac operator $\feyn D + (\mu/N_c)\gamma_0$:
\begin{equation}
D=\left(
\begin{array}{cc}
0  &  iX + iC\\
iX^{\dag} + iC & 0
\end{array}
\right) \ .
\end{equation}
The random matrix $X$ has dimension $N\times N$.  The total dimension of 
$D$ is $2N$.  This is the number of small eigenvalues, which is proportional 
to $V_4$.   In QCD we expect $N$ to be approximately
equal to the typical number of instantons (or anti-instantons)
in $V_4$; therefore, $N/V_4 \approx n_{\rm inst} \approx 0.5\, {\rm
fm}^{-4}$ \cite{Sh82}. The matrix $C$ is 
deterministic and describes the effects of temperature and chemical 
potential.  In the simplest (and original) $T \not= 0$, $\mu=0$ model 
\cite{JaVe96}, the choice $C=\pi T$ describes the effect of the smallest 
Matsubara frequency.  As noted in ref.\,\cite{WeSc96}, it is possible to 
simulate the effects of the eigenvalue correlations induced by the pairing 
of instantons and anti-instantons into molecules by choosing a more general 
form for the diagonal matrix $C$ with elements, $C_k$, which are (increasing) 
functions of $T$.  In the $T=0$, $\mu \not= 0$ model of ref.\,\cite{St96b},
$C=\mu/(iN_c)$ describes the effect of the chemical potential. In this paper, 
we consider the more general case $T \not= 0$, $\mu \not= 0$. Although we 
do not know the detailed dependence of the elements of $C$ on $T$ and $\mu$,
we understand that $T$ primarily affects the real (i.e., hermitian) part of
$C$ and $\mu$ affects the imaginary (i.e., antihermitian) part.  We shall 
adopt the following approximate form for this dependence, $C_k=a\pi T+
b\mu/(iN_c)$ for one half of eigenvalues and 
$C_k=-a\pi T+b\mu/(iN_c)$ for the other half with $a$ and $b$
dimensionless parameters.\footnote{These 
parameters are intended to reflect the degree of overlap and correlation 
between instantons and anti-instantons. One can therefore anticipate that 
$a$ and $b$ are smaller than 1.  We shall estimate the values of $a$ and 
$b$ below.}  This form accounts for the fact that there are two smallest
Matsubara frequencies\footnote{Alternatively, this form preserves the 
relation $\langle \gamma_0 D \rangle=0$ at $\mu=0$.}, $+\pi T$ and
$-\pi T$.  Such a linear ansatz for $C$ is certainly
very naive, but in this paper we decided not to try and refine it.
This form reflects sufficiently well our understanding of
the properties of $C$.

The chiral condensate is calculated as:
\begin{equation}\label{pbp=zrm}
\langle\bar\psi\psi\rangle = {1\over N_f V_4}{\partial 
\ln Z_{\rm RM}\over\partial m}.
\end{equation}
Current algebra fixes the value of $\langle {\bar \psi} \psi \rangle_0 
\approx 2\, {\rm fm}^{-3}$ at $T=\mu=m=0$. The only dimensionful parameter 
remaining in the partition function, $Z_{\rm RM}$, is the variance 
of the random matrix, $\sigma$.  Thus, 
\begin{equation}\label{pbp=ninst}
\langle\bar\psi\psi\rangle_0 = {\rm const} {N\over V_4\sigma}
\approx{\rm const} {n_{\rm inst}\over \sigma} \ .
\end{equation}
The dimensionless constant will be found below and is equal to 2.
This fixes the value of $\sigma\approx 0.5\,{\rm fm}^{-1}\approx 100$ MeV.
It is convenient to use $\sigma$ as a unit of mass in the model
and also absorb the coefficients $\pi a$ and $b/N_c$ into
$T$ and $\mu$. In other words, we measure $m$ in units of $\sigma$,
$T$ in units of $\sigma/(\pi a)$ and $\mu$ in units of $\sigma N_c/b$.

The $N \to \infty$ (i.e., thermodynamic) limit of the partition function, 
eqn.\,(\ref{ZRM}), can be found in the now-standard way 
\cite{JaVe96,St96a,WeSc96}. Performing the Gaussian integration over $X$
and introducing auxiliary $N_f\times N_f$ matrices 
$\phi$, one can rewrite the partition function in the form
\begin{eqnarray}\label{zphi}
&Z_{\rm RM} &
= \int {\cal D}\phi \exp[-N{\rm Tr}(\phi\phi^{\dag})]
\, {\det}^{N/2} \left(
\begin{array}{cc}
\phi + m & \mu + iT\\
\mu + iT & \phi^\dag + m
\end{array}
\right)
{\det}^{N/2} \left(
\begin{array}{cc}
\phi + m & \mu - iT\\
\mu - iT & \phi^\dag + m
\end{array}
\right)
\nonumber\\
&=&
\int {\cal D}\phi \exp[-N\Omega(\phi)],
\end{eqnarray}
where 
\begin{equation}
\Omega(\phi)={\rm Tr}[\phi\phi^\dag 
- \frac12\ln\{
[(\phi+m)(\phi^\dag+m)- (\mu+iT)^2]\cdot
[(\phi+m)(\phi^\dag+m)- (\mu-iT)^2]
\}].
\end{equation}
The integration in eqn.\,(\ref{zphi}) is performed over
$2\times N_f\times N_f$ variables which are the real and imaginary
parts of the elements of the complex matrix $\phi$.
In the limit $N\to\infty$ this integral is 
determined by a saddle point of the integrand or, alternatively, the 
minimum of $\Omega(\phi)$:
\begin{equation}\label{lnz=v}
\lim_{N\to\infty} {1\over N}\ln Z_{\rm RM} = - \min_\phi \Omega(\phi).
\end{equation}
The function $\Omega(\phi)$ is an effective potential for the degrees of 
freedom describing the dynamics of the chiral phase transition.  One can 
see that the value of $\phi$ at the minimum, i.e., the equilibrium value 
$\langle \phi \rangle$, gives us the value of the chiral condensate, 
eqn.\,(\ref{pbp=zrm}):
\begin{equation}\label{pbp=phi}
\langle\bar\psi\psi\rangle = 
{1\over N_f V_4}{N\over\sigma}\,2{\rm Re}{\rm Tr}\langle\phi\rangle,
\end{equation}
(cf. eqn.\,(\ref{pbp=ninst})).

For real $m$, it is reasonable to expect that the minimum occurs when $\phi$
is a real matrix proportional to a unit matrix. 
With this assumption, we need to find only
one real parameter, $\phi$, which minimizes the potential:
\begin{equation}
\Omega = N_f\left[\phi^2 - \frac12\ln\left\{
[(\phi+m)^2- (\mu+iT)^2]\cdot
[(\phi+m)^2- (\mu-iT)^2]\right\}\right] \ .
\end{equation}
This is not a simple $\phi^6$ potential, but
it has remarkably similar properties. In particular, the condition
$\partial \Omega/\partial \phi=0$ gives a fifth-order polynomial equation
in $\phi$. Let us first consider the symmetry plane $m=0$.
Then the equation $\partial \Omega/\partial \phi=0$ always has one trivial 
root, $\phi=0$.
The remaining four roots are the solutions of a quartic equation,
which has the form of a quadratic equation in $\phi^2$:
\begin{equation}\label{biquad}
\phi^4 - 2\left(\mu^2-T^2+\frac12\right)\phi^2 + (\mu^2+T^2)^2 
+ \mu^2 - T^2 = 0.
\end{equation}
Above the second-order line, i.e., in the high temperature phase, 
eqn.\,(\ref{biquad}) does not have real roots 
(since $\phi^2 < 0$ for 
each pair of roots). This corresponds to the fact that the potential
$\Omega$ has only one minimum at $\phi=0$ (i.e., the trivial root). 
On the second-order line,
a pair of roots of eqn.\,(\ref{biquad}) becomes zero, i.e., the
potential is $\Omega\sim\phi^4$ near the origin. This means that
on the second-order line,
\begin{equation}\label{phi0}
(\mu^2+T^2)^2 + \mu^2 - T^2 = 0 \ .
\end{equation}
The second-order line ends when the remaining pair of roots also becomes
zero, i.e., the potential becomes $\Omega\sim\phi^6$ near the origin.
This happens when
\begin{equation}\label{phi2}
\mu^2-T^2+\frac12 = 0
\end{equation}
on the second-order line. The condition, eqn.\,(\ref{phi2}) together with
eqn.\,(\ref{phi0}) determines the location of the tricritical point
in the $T \mu$ plane:
\begin{equation}
T_3 = \frac12\sqrt{\sqrt2+1} \approx 0.776
\qquad\mbox{and}\qquad
\mu_3 = \frac12\sqrt{\sqrt2-1} \approx 0.322 \ .
\end{equation}

The equation for the triple line is obtained from the requirement 
that the depth of the minima in $\Omega$ at $\phi$ given by the pair 
of solutions of eqn.\,(\ref{biquad}) (farthest from the origin) 
should coincide with the depth at the origin $\phi=0$. The equation for 
the triple line is therefore
\begin{equation}
\mu^2 - T^2 + \frac12 + \frac12\sqrt{1-16\mu^2T^2}
- \frac12 \ln\left(1+\sqrt{1-16\mu^2T^2}\over2\right) + \ln(\mu^2+T^2) = 0.
\end{equation}
In particular, when $T=0$, we obtain the elementary equation $\mu^2 + 1 
+ \ln\mu^2 = 0$ whose solution is $\mu \approx 0.528$ \cite{St96b}. 
This is the value of $\mu_1$.  Setting $\mu=0$ in the equation for 
the second-order line, eqn.\,(\ref{phi0}), we find $T_c=1$ \cite{JaVe96}.

We recall that the units of $T$ and $\mu$ depend on the unknown
dimensionless parameters $a$ and $b$.  However, these unknown factors 
cancel from the ratios
\begin{equation}
{T_3\over T_c} \approx 0.78; \qquad {\mu_3\over\mu_1} \approx 0.61. 
\end{equation}
Taking $T_c=160$ MeV and $\mu_1=1200$ MeV, we find that $T_3 \approx 
120$ MeV and $\mu_3\approx 700$ MeV.\footnote{With these values for $T_c$ 
and $\mu_1$, the parameters $a$ and $b$ have the values $a \approx  0.2$ 
and $b \approx 0.13$.}

Note that the second-order line, eqn.\,(\ref{phi0}), marks the location
of the points on the phase diagram where the symmetric minimum
$\phi=0$ disappears, i.e., turns into a maximum. Continuing this
line below point $T_3$, we obtain the location of spinodal points.
In the region between this line of spinodal points and the
first-order phase transition line, the chirally symmetric phase 
$\phi=0$ can exist as a metastable state. Such a state can be reached by
supercooling, and it is unstable towards the nucleation of bubbles of
the broken phase $\phi\ne0$. A similar line with equation $4T\mu=1$,
the superheating line, together with the supercooling line,
eqn.\,(\ref{phi0}), bound the region around the first-order phase
transition line where the potential $\Omega(\phi)$ has 3 minima. All
these lines meet at the tricritical point.

Away from the symmetry plane $m=0$, the expressions for the wing
surfaces and the wing lines become rather lengthy and will be presented
elsewhere.  The principle, however, remains simple.  The minima of the
potential, $\Omega$, satisfy the equation $\partial \Omega/\partial \phi=0$
and are given by (three out of five) roots of a fifth-order polynomial.
On the wing surface, the depth $\Omega(\phi)$ in a pair of adjacent minima
is the same. On the wing line, the two adjacent minima fuse into one.
In terms of the roots of the polynomial, three roots coincide (two minima 
and one maximum).  In other words, the potential is $\Omega \sim 
(\phi-\langle\phi\rangle)^4$ on the wing line and near the minimum.

\begin{figure}[htb]
\setlength{\unitlength}{2.4in}
\centerline{\psfig{file=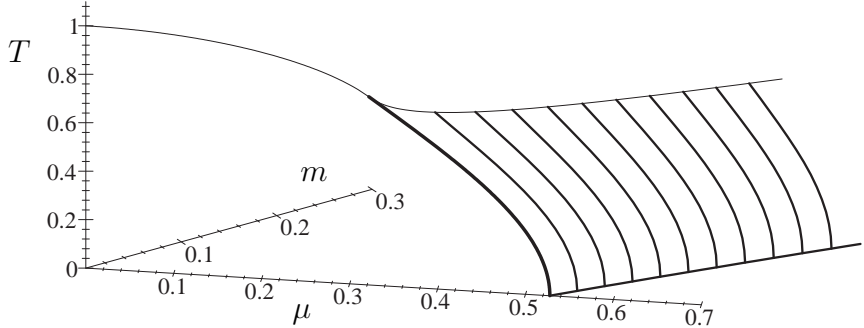,width=2\unitlength}}
\vspace{-\unitlength}
\begin{picture}(2,1)
\put(.34,.62){ $T$}
\put(.96,.06){ $\mu$}
\put(1.00,.36){$m$}
\end{picture}
\caption[]{\small Phase diagram of QCD with two light flavors
of mass $m$ as calculated from the random matrix model. 
The almost parallel curves on the wing surface
are cross sections of this surface with $m=$const planes.
The units of $m$ are $\sigma\approx 100$ MeV, of $T$ are
$T_c\approx 160$ MeV, of $\mu$ are $\mu_1/0.53\approx 2300$ MeV,
with the choices of $T_c$ and $\mu_1$ from the text.
}
\label{fig:3dRM}
\end{figure}

The resulting phase diagram is plotted in Fig.~\ref{fig:3dRM}.  One
can see, as expected from mean field theory near the tricritical
point, that the wing lines together with the triple line approach the
tricritical point with the same slope as the second-order line but from the
other side.  The critical exponents near the tricritical point as 
given by the random matrix model can be also seen to coincide, as
expected, with the mean field exponents of eqns.\,(\ref{deltapbp}),
(\ref{deltan}), and (\ref{pbp1/4}).

\begin{figure}[hbt]
\centerline{
\psfig{file=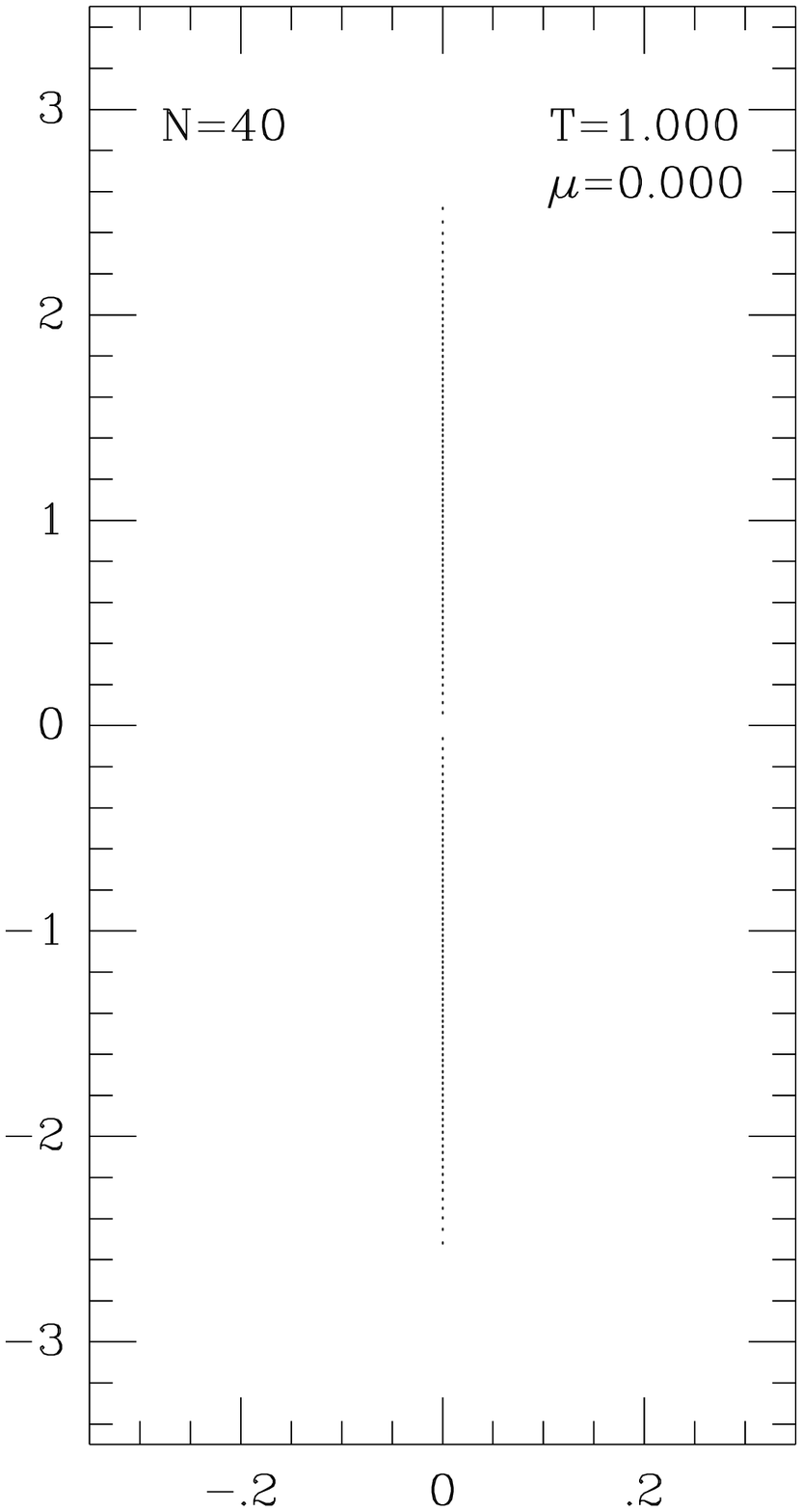,width=1.6in}
\psfig{file=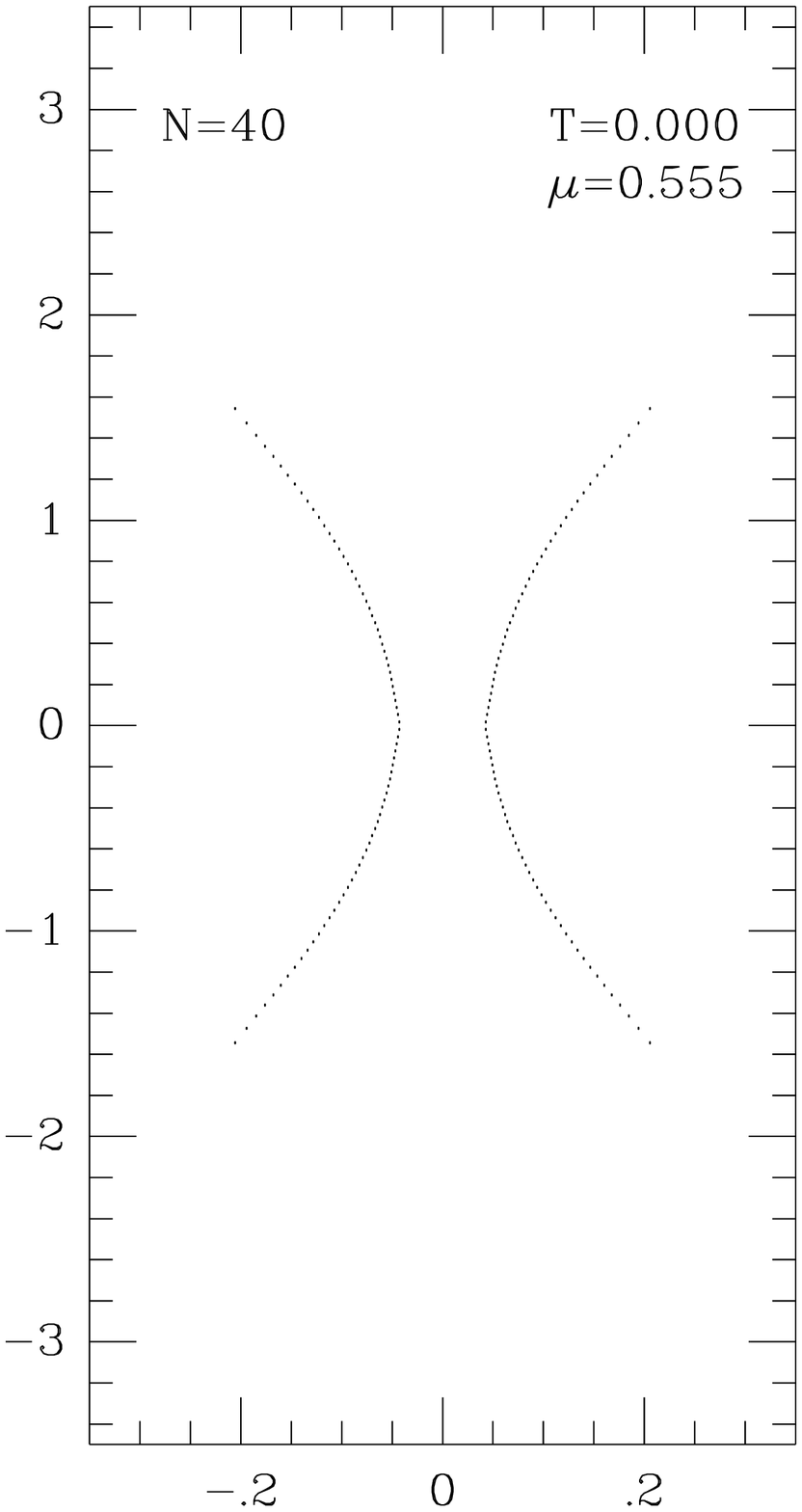,width=1.6in}
\psfig{file=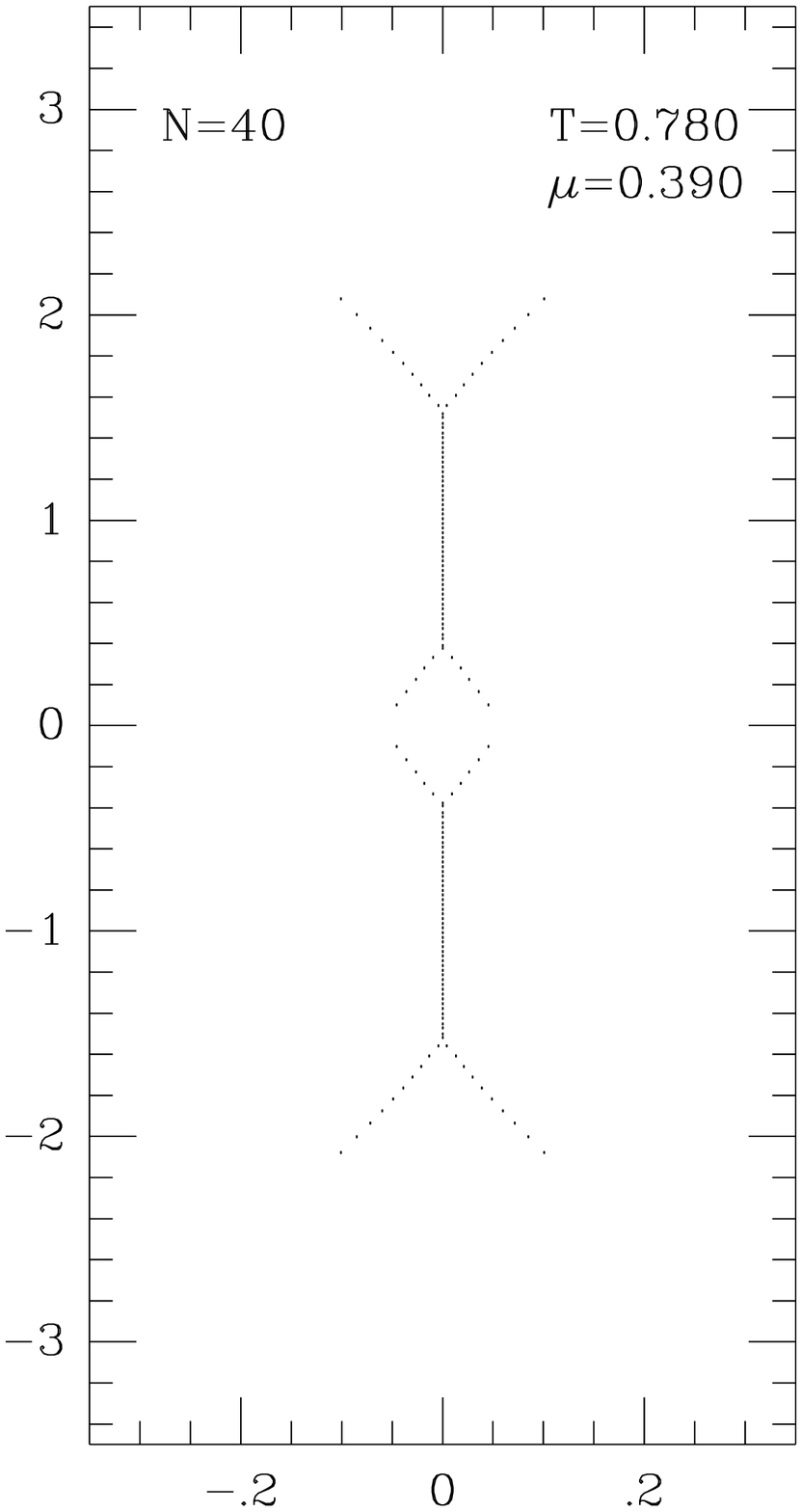,width=1.6in}
}
\centerline{(a)\hfil(b)\hfil(c)}
\caption[]{\small Zeros of the partition function of a finite size $N$ random
matrix model (\ref{ZRM}) in the complex $m$ plane calculated
numerically at different values of $T$ and $\mu$. The calculation
is done for $N_f=1$, but the $N\to\infty$ limit is $N_f$ independent.
The density of points is proportional to the strength of the
cut (discontinuity in $\langle\bar\psi\psi\rangle$) in the
$N\to\infty$ limit.
}
\label{fig:zeros}
\end{figure}

\begin{figure}[hbt]
\setlength{\unitlength}{3.2in}
\centerline{
\psfig{file=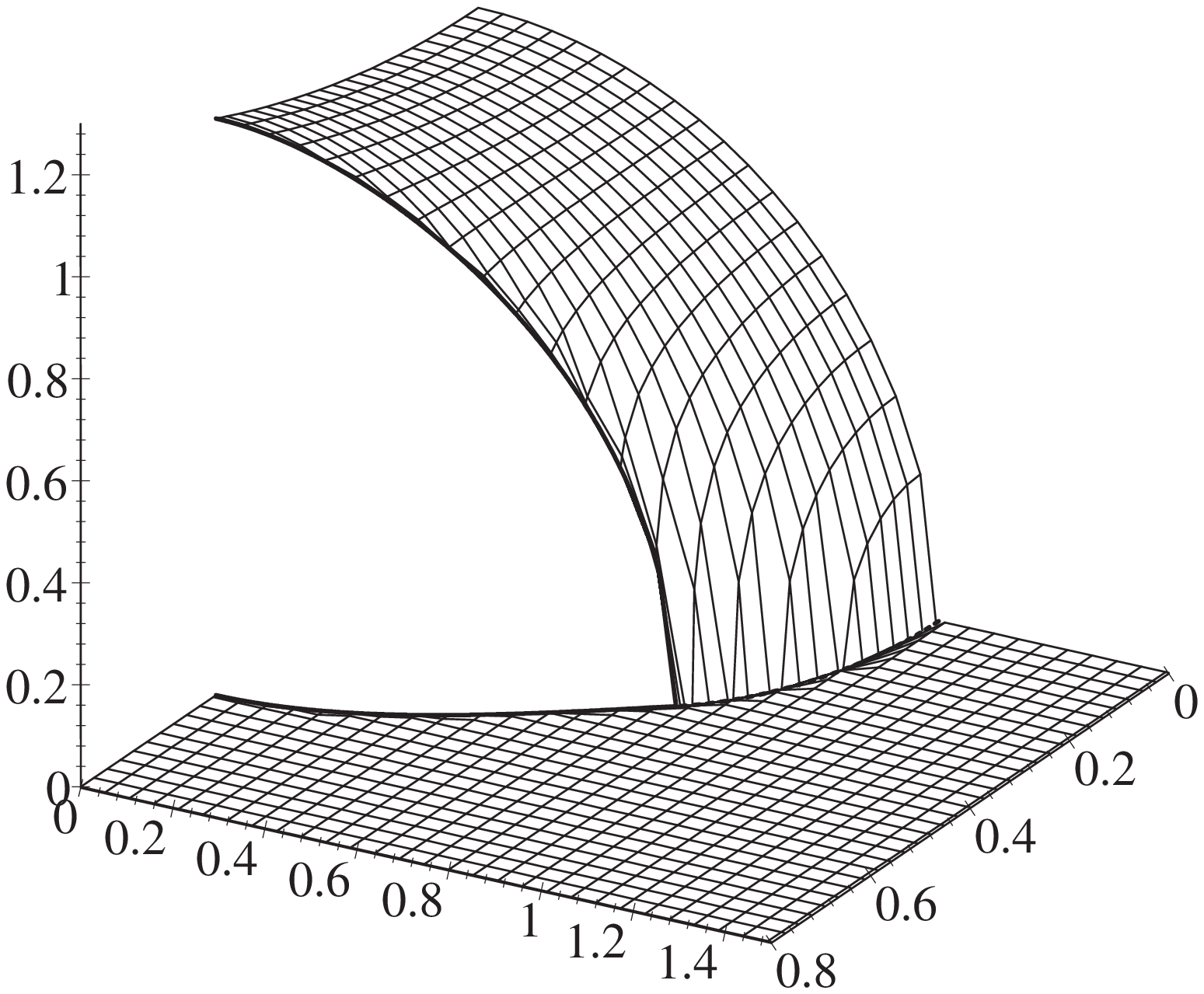,width=\unitlength}}
\vspace{-\unitlength}
\centerline{
\begin{picture}(1,1)
\put(.23,.03){ $T$}
\put(.78,.06){ $\mu$}
\put(-.1,.60){$\langle \bar{\psi} \psi \rangle$}
\end{picture}}
\caption[]{\small The chiral condensate $\langle\bar\psi\psi\rangle$ (in units
of $\langle\bar\psi\psi\rangle_0\approx2\,{\rm fm}^{-3}$) as a
function of $T$ and $\mu$ in the random matrix model. The units
of $T$ and $\mu$ are as in Fig.~\ref{fig:3dRM}.
}
\label{fig:bread}
\end{figure}

 From the random matrix model, we also learn how the zeros of the
partition function in the complex $m$ plane evolve with changes in 
temperature and chemical potential.  A few typical cases are illustrated 
in Fig.~\ref{fig:zeros}.  At zero $T$ and $\mu$, the zeros form a 
cut (in the $N\to\infty$ limit) along imaginary axis.  Raising the temperature 
pushes the zeros away from the origin along the imaginary axis until the 
density at the origin vanishes (continuously), the cut breaks in two, 
as in Fig. \ref{fig:zeros}a, and chiral symmetry is restored (cf. 
eqn.\,(\ref{YL})).  The chemical potential pushes the zeros away from 
the origin in the direction of the real axis until the cut splits in two, as 
illustrated in Fig. \ref{fig:zeros}b.  
Note that the density $\rho(0)$ is finite just 
before the split.  Therefore, the transition is of first order.  Near the 
tricritical point, the split in the direction of the real axis (due to 
the chemical potential) occurs at the same time that the density $\rho(0)$ 
vanishes (due to the effects of the temperature).  This is illustrated 
in Fig. \ref{fig:zeros}c. The resulting dependence of
$\langle\bar\psi\psi\rangle$ on $T$ and $\mu$ is shown in
Fig. \ref{fig:bread}.

A comment should be added regarding the calculation of the baryon 
density, $n$, in the random matrix model. The value of $(1/V_4) 
\partial \ln Z_{\rm RM}/ \partial \mu$ does not represent the 
complete baryon density.  The reason is that $\ln Z_{\rm RM}$ contains 
only contributions from the soft modes of the condensate, $\phi \sim 
{\bar \psi} \psi$.  Further dependence on $\mu$ is contained in the 
contributions to $\ln Z$ from other degrees of freedom.  In the effective
Landau-Ginzburg theory, such additional contributions are embedded in 
the term $\Omega_0(T,\mu)$, eqn.\,(\ref{LG}).  However, the {\em singular} 
behavior of the system is exclusively due to the soft modes of the 
condensate, i.e., terms involving $\phi$ in $\Omega_{\rm eff}$.  Thus, 
it is legitimate to calculate singular properties of $n$, such as 
$\Delta n$, near a (tri)critical point using $(1/V_4)\partial \ln 
Z_{\rm RM}/\partial\mu$.  At $T=0$, for example, we find:
\begin{equation}
\Delta n V_4 = N \left[
\left(\partial\Omega\over\partial\mu\right)_{\phi=0} -
\left(\partial\Omega\over\partial\mu\right)_{\phi=\sqrt{1+\mu_1^2}}\right] =
NN_f\left({2\over\mu_1}+2\mu_1\right)\approx 5NN_f,
\end{equation}
in units of $1/\mu$, which is $b/N_c\Sigma$.  With our previous choice of 
$\mu_1 = 1200$ MeV (i.e., $b\approx 0.13$), we find that $\Delta n \approx 
0.4\, {\rm fm}^{-3} \approx 2.5 \, n_0$, which seems a reasonable estimate.

\section{Discussion and summary}

\begin{figure}
\epsfysize 2.5in\centerline{\epsfbox{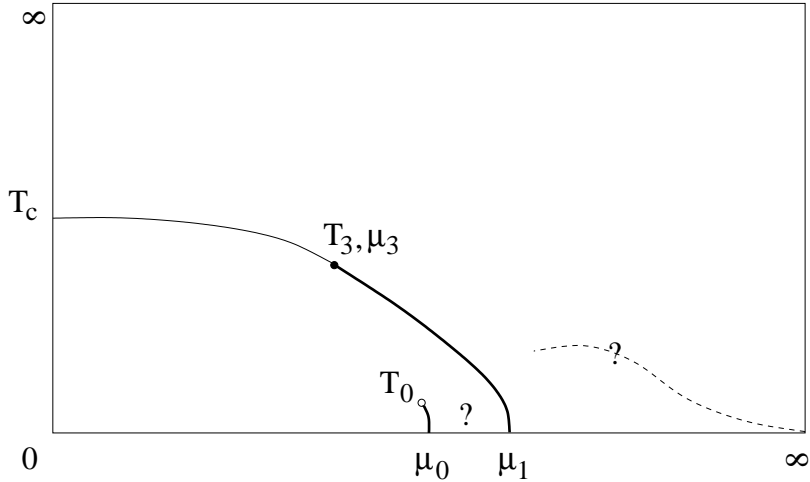}}
\caption[]{\small A schematic phase diagram of QCD with 2 massless
quark flavors. Other phase transition lines are possible, for example,
in the low temperature region to the right of $\mu_0$. Another
example is a transition associated with color superconductivity
plotted as a dashed line. Thicker lines are first-order phase
transitions. The $T_c-T_3$ line is a second-order phase transition.
The tricritical point is at $T_3$, $\mu_3$  and the critical point
of the nuclear matter liquid-gas transition is at $T_0$.}
\label{fig:pd}
\end{figure}

In this paper, we have presented an analysis (qualitative and, in some 
cases, quantitative) of the salient features of the phase diagram of QCD 
with two light or massless quark flavors at finite temperature and baryon 
chemical potential.  The most important features of this phase diagram are 
summarized in Fig.~\ref{fig:pd}. The phase diagram can certainly have
a much richer structure. The phase transitions shown there are 
distinguished by the fact that a good order parameter can be associated 
with each of them.  Here the term ``good order parameter'' implies the 
existence of some quantity whose expectation value is identically
zero in some finite region of parameter space or in one phase and is 
some function of parameters in the other phase.  Two such phases must 
be separated by a nonanalytic boundary, i.e. a phase transition.  
What is crucial here is the identical vanishing of an order parameter or its 
strict
independence of the parameters of the theory.  Usually, this is ensured 
by the existence of some symmetry with respect to which this order 
parameter transforms nontrivially.

For the chiral phase transition, a good order parameter is the value of
$\langle {\bar \psi} \psi \rangle$, which spontaneously
breaks the global SU(2)$_L \times$ SU(2)$_R$ chiral 
symmetry to $SU(2)_V$.  
Hence, the phase with $\langle {\bar \psi} \psi \rangle = 0$ and the 
phase with $\langle {\bar \psi} \psi \rangle \not= 0$ cannot be connected
without crossing a phase transition line in the $T \mu$ plane.

The transition from $n \equiv 0$ to $n \not= 0$ along the $T=0$ line 
provides another example of a phase transition associated with a good order 
parameter.  The phases $n \equiv 0$ and $n \not= 0$ cannot be analytically
connected; i.e. one must pass through a nonanalytic boundary 
(phase transition)
when passing from one to the other.  Since this is a first-order phase 
transition, continuity requires that there is also a first-order  
transition line for some $T \not= 0$.  This line can, however, terminate 
since $n$ is no longer a good order parameter when $T \not= 0$.

The existence of a good order parameter is a sufficient but
not a necessary condition for a phase transition.
Other phase transition lines 
associated with more subtle phenomena are also possible. 
One interesting
example, which attracted attention recently, is the transition
associated with color superconductivity. The existence of a
color superconducting phase was first argued for
by Bailin and Love on the grounds that one gluon exchange
is attractive in the color antitriplet quark-quark channel
\cite{BaLo84}.
This means that the Fermi surface at very high $\mu$ becomes
unstable and forms a gap. 
This phenomenon was recently reanalyzed
using other methods at moderate values of $\mu$ with the conclusion
that the effect is enhanced by instanton-induced
interactions \cite{AlRa97,RaSc97}.
This means that another finite-$T$ transition, which stretches
all the way to $\mu=\infty$, may be present on the phase diagram,
Fig.~\ref{fig:pd}.

This transition, unlike the chiral phase transition and the nuclear matter 
liquid-gas transition (at $T=0$), does not seem to have a good order 
parameter associated with it. In particular, the diquark condensate, 
$\langle \psi \psi \rangle$, is not gauge invariant.  The dynamical 
mechanism responsible for the binding of diquark pairs is certainly 
operative at low temperatures, but the absence of a good order parameter 
does not allow us to assert that the temperature induced transition 
associated with the breaking of diquark pairs must always (i.e., at all 
$\mu$) proceed through a thermodynamic singularity rather than a smooth 
analytic crossover.

\begin{figure}
\centerline{\epsfysize 2.5in \epsfbox{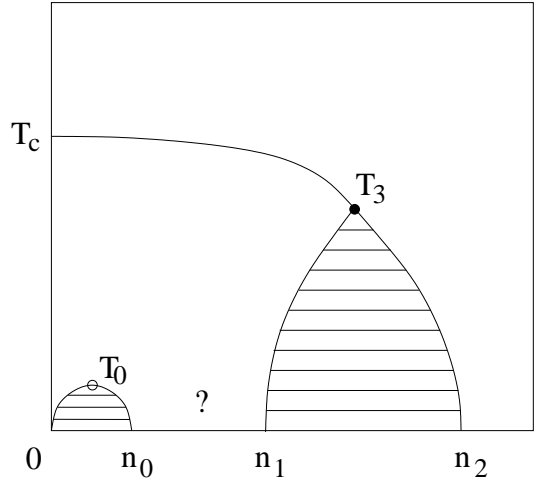}}
\caption[]{\small The phase diagram of Fig.~\ref{fig:pd} shown as a 
function of density and temperature.  The dashed line of the
color superconductivity transition is not drawn.  Horizontal lines 
connect points corresponding to densities of phases on two sides of the
first-order line (i.e., the coexistence curve) of Fig.~\ref{fig:pd}.
The points $n_i$ on the
$T=0$ line are the same as on Fig.~\ref{fig:nmu}a.}
\label{fig:pdnT}
\end{figure}

For some purposes, it is more natural to study the phase diagram in the 
space of density and temperature.  The phase diagram of Fig.~\ref{fig:pd} 
can be converted into such a space and is shown in Fig.~\ref{fig:pdnT}.  
A typical feature is that the first-order phase transition line from the 
$T \mu$ plane now appears as a region of phase coexistence in 
Fig.~\ref{fig:pdnT}.  In equilibrium, the values of density and temperature 
inside this region can be achieved only through an {\em inhomogeneous\/} 
mixture of two phases with different densities but the same $T$ and $\mu$.
These densities are indicated by the ends of the horizontal lines drawn 
in the phase coexistence region.

The most interesting feature of the phase diagram of Figs.~\ref{fig:pd} and 
\ref{fig:pdnT} is the presence of a tricritical point.  Because of the fact
that the critical dimensionality for such a point is equal to 3, critical 
behavior near this point is given by mean field theory plus
logarithmic corrections.  In particular, a 
simple random matrix model predicts the correct algebraic critical exponents.

The tricritical point lies in the region expected to be probed by heavy ion 
collision experiments.  It would be interesting to find an experimental 
signature for such a point. Since quark masses are not precisely zero, we 
should consider a slice of a three-dimensional phase diagram, 
Fig.~\ref{fig:3dRM}, with $m \ne 0$.  A qualitative difference between 
the phase diagram for $m \not= 0$ and that for $m=0$ is the absence of 
the second-order phase transition line associated with the restoration of 
chiral symmetry.  This symmetry is explicitly broken for $m \ne 0$.  However, 
continuity from $m=0$ ensures that the first-order finite density transition 
is still present at $m\ne0$. 
This transition line is terminated by an ordinary critical 
point.  Criticality at this point is not associated with chiral symmetry 
restoration, and excitations with the quantum numbers of pions do not
become massless there.  Criticality at this point is associated with the 
fact that a correlation length in the channel with the quantum numbers of the
sigma meson becomes infinite. (Hence, it is plausible to infer that
this point has the critical behavior of the three-dimensional Ising
model.) Possible 
experimental signatures of this phenomenon are under investigation.

\subsection*{Acknowledgments}

Discussions with G. Brown, R. Pisarski, M. Prakash, K. Rajagopal, 
and E. Shuryak are greatly appreciated. This work was supported in
part by  DOE grant DE-FG-88ER40388 and NSF grant PHY-97-22101.

\subsection*{Note added}

After this work was completed, a paper \cite{BeRa98} appeared
which addresses similar questions in the context of a Nambu-Jona-Lasinio
model for color superconductivity. The results of \cite{BeRa98}
agree with and complement our findings.

\end{document}